 \documentclass[final,onecolumn,3p,times,sort&compress,12pt]{elsarticle}

\usepackage{graphicx}
\usepackage{dcolumn}
\usepackage{bm}
\usepackage{epstopdf}
\usepackage{bm,hyperref,color,breakurl}

	\usepackage{fancyheadings}
\renewcommand{\thispagestyle}[1]{} 

\begin{document}

\begin{frontmatter}



\title{Hubbard pair cluster in the external fields. Studies of the polarization and susceptibility}


\author{T. Balcerzak}
\ead{t\_balcerzak@uni.lodz.pl}
\author{K. Sza{\l}owski\corref{cor1}}
\ead{kszalowski@uni.lodz.pl}
\address{Department of Solid State Physics, Faculty of Physics and Applied Informatics,\\
University of \L\'{o}d\'{z}, ulica Pomorska 149/153, 90-236 \L\'{o}d\'{z}, Poland}

 \cortext[cor1]{Corresponding author. E−mail address:
kszalowski@uni.lodz.pl}

\date{\today}

\begin{abstract}
The electric and magnetic polarizations as well as the electric and magnetic susceptibilities of the Hubbard pair-cluster embedded in the external fields were studied by the exact method. Based on the grand canonical ensemble for open system, the numerical calculations were performed for the electron concentration corresponding to the half-filling case. It has been found that the electric and magnetic properties are strictly interrelated, what constitutes a manifestation of a magnetoelectric effect, and the detailed explanation of such behaviour was given.
In particular, near the ground state where the transitions are induced by the external fields, discontinuous changes of the studied quantities have been found.  They have been associated with the occurrence of the singlet-triplet transitions. An anomalous behaviour of the electric and magnetic polarizations as a function of the temperature, occurring below the critical magnetic field, was illustrated. In the presence of the competing electric and magnetic fields, the influence of Coulombic repulsion on the studied properties was discussed.\\
\end{abstract}


\end{frontmatter}

\section{Introduction}

The Hubbard model, since its formulation \cite{Anderson1959, Hubbard1963, Gutzwiller1963, Kanamori1963}, has been intensively studied by the solid state physicists.
Being the first model capable to describe the metal-insulator (Mott) transition, it has also been studied in relation to such problems as the magnetic phase transitions, high-temperature superconductivity, optical lattices and graphene properties \cite{Chen1979, Ho1979, Hirsch1980, Robaszkiewicz1981a, Robaszkiewicz1981, Hirsch1985, Lieb1968, Lieb2003,  Hirsch1989, Sorella1992, Pelizzola1993, Janis1993, Staudt2000, Peres2004, Kent2005, Zaleski2008, 
Joura2015, Li2015, McKenzie1998, Fuchs2011, Rohringer2011,  Yamada2014, Kozik2013,  Karchev2013, Claveau2014, Lieb1989, Shastry1986, Su1992, Mancini2009, Tocchio2013, Dang2015, Mermin1966, Nolting2009a, Dombrowsky1996, Feldner2010, Weymann2015, Chao1977, Yosida1998, Tasaki1998, Micnas1990, Georges1996, Hirschmeier2015, Mielke2015}.

Despite numerous theoretical efforts, the rigorous solutions to the Hubbard model for infinite systems have been obtained in very few cases only. The exact results include, for instance, the solution for one-dimensional (1D) system \cite{Lieb1968, Lieb2003, Shastry1986, Su1992} as well as several rigorous theorems, to mention Mermin-Wagner theorem in 2D systems \cite{Mermin1966, Nolting2009a, Tasaki1998} or Lieb theorems for the ground state \cite{Lieb1989}.

At the same time, it has been noticed that the exact solutions to the model can be obtained for small clusters, consisting of several lattice sites \cite{Schumann2008,Schumann2007, Cisarova2014, Cencarikova2016, Galisova2015, Galisova2015a, Harris1967, Silantev2015, Hasegawa2005, Hasegawa2011, Spalek1979, Longhi2011, Juliano2016, Kozlov1996, Alvarez-Fernandez2002, Avella2003, Szalowski2015,  Szalowski2017, Balcerzak2017, Balcerzak2018,Wortis2011,Wortis2017, Cheng1976, Ohta1994, Noce1997, Iglesias1997,Amendola2015, Becca2000, Ricardo-Chavez2001, Yang2009, Ovchinnikova2013, Carrascal2015, Fuks2014, Kamil2016, Souza2016, Balasubramanian2017}. Intensive investigations of such systems have been carried out both from the point of view of static properties 
\cite{Cheng1976, Spalek1979, Ohta1994, Noce1997, Amendola2015,Becca2000, Ricardo-Chavez2001, Alvarez-Fernandez2002, Schumann2007, Schumann2008, Yang2009, Schumann2010, Ovchinnikova2013, Cisarova2014, Hancock2002,Hancock2005,Hancock2014,Galisova2015, Szalowski2015, Carrascal2015, Kamil2016, Souza2016, Balcerzak2017, Balcerzak2018,Ullrich2018}, as well as for dynamical description \cite{Harris1967, Kozlov1996, Longhi2011, Fuks2014, Wortis2017, Balasubramanian2017}.
In case of very small atomic clusters, exact results for the Hubbard model have been obtained by analytical methods \cite{Harris1967, Cheng1976, Noce1997,Iglesias1997, Amendola2015,Alvarez-Fernandez2002, Schumann2007, Schumann2008, Schumann2010, Ovchinnikova2013, Cisarova2014, Galisova2015, Carrascal2015, Kamil2016, Balcerzak2017, Balcerzak2018, Balasubramanian2017}. However, for larger clusters the numerical techniques turned out to be indispensable \cite{Spalek1979, Ohta1994, Becca2000, Ricardo-Chavez2001, Yang2009, Szalowski2015, Souza2016, Szalowski2017}.
It is worth mentioning that theoretical studies of finite clusters are becoming increasingly important for the development of experimental nanophysics and nanotechnology.

The simplest system, for which the Hubbard model can be solved analytically, is a two-site atomic cluster (dimer). Despite many theoretical works, the system has not been fully examined yet. For instance, this concerns the case when the two-site cluster is simultaneously embedded in two external fields: magnetic and electric one, and is able to exchange the electrons with its environment. Such a system can model a physical situation where the atomic dimer is deposited on the surface and interacts both with the surface and the external fields. The influence of the electric field, acting as a control factor, on the magnetic properties of the cluster constitutes a manifestation of magnetoelectric effect and is very interesting from the point of view of possible application, for instance, in spintronics and/or memory devices. Some examples can be recalled here, mainly to mention the molecular dimer systems. Among them molecular mixed-valence dimers \cite{Soncini2010,Bosch-Serrano2012,Suzuki2014,Palii2014,Dudnik2015,Pansini2018} or $\kappa$-(BEDT-TTF) \cite{Naka2016} focus particular attention and appear highly promising; however, also some non-molecular systems such as dimers on graphene surface \cite{Hu2014} also attract the interest.

The theoretical studies of two-atomic Hubbard cluster, treated as a thermodynamic open system and placed simultaneously in two external fields, have been initiated in the papers \cite{Balcerzak2017} and \cite{Balcerzak2018}. In Ref.~\cite{Balcerzak2017} the main formalism has been presented and thorough investigations of the chemical potential have been carried out. On this basis, in the paper \cite{Balcerzak2018} the studies have been extended to magnetic properties, concentrating mainly on the phase diagrams, cluster magnetization, spin-spin correlation functions and mean hopping energy. 

The aim of the present work is a continuation of these studies, basing on the formalism developed in Ref.~\cite{Balcerzak2017}, towards elucidation of interesting interrelations between magnetic and electric properties for the Hubbard dimer  exhibiting a non-trivial magnetoelectric behaviour. In particular, the electric polarization of the cluster, as well as the electric susceptibility in the external fields will be studied. Simultaneously, the magnetic polarization and magnetic susceptibility will be analysed. A comparison of the magnetic and electric properties will be done, which seems interesting not only from the purely theoretical point of view for this model. In our opinion, the magnetoelectric correlations existing between the described measurable quantities may be also of practical interest, giving the possibility of controlling the magnetic state of the cluster by the electric potential.

The paper is organized as follows: In the theoretical Section \ref{theory} the model is briefly presented and the basic quantities, important for numerical calculations, are defined. In the successive Section \ref{num} the numerical results are illustrated in figures and discussed. 
An extensive comparison of magnetic and electric properties is performed there.
The last Section \ref{conc} is devoted to a brief summary of the results and concluding remarks. The \ref{app} collects the expressions for the eigenenergies corresponding to the quantum states with two electrons per dimer and shows the behaviour of these states as a function of the electric and magnetic field.

\section{\label{theory}Theoretical model}

The Hamiltonian of the Hubbard pair-cluster (dimer) consisting of $(a,b)$ atoms and interacting with the external fields is assumed in the form:
\begin {eqnarray}
\mathcal{H}_{a,b}&=&-t\sum_{\sigma=\uparrow,\downarrow}\left( c_{a,\sigma}^+c_{b,\sigma}+c_{b,\sigma}^+c_{a,\sigma} \right)+U\left(n_{a,\uparrow}n_{a,\downarrow}+n_{b,\uparrow}n_{b,\downarrow}\right)\nonumber\\
&&-H\left(S_a^z+S_b^z\right) -V\left(n_{a}-n_{b}\right),
\label{eq1}
\end {eqnarray}
where $t>0$ is the hopping integral and $U\ge 0$ is the on-site Coulomb repulsion parameter. The symbol $H=-g\mu_{\rm B}H^z$ stands for an external uniform magnetic field $H^z$ oriented along $z$-direction. The term with $V$ introduces the potential energy of the atoms $a$ and $b$ in the electric field. For such potential distribution the external electric field $E$ is oriented along the pair and is equal to $E=2V/\left(|e|d\right)$ with $d$ being the interatomic distance, whereas $e$ is the electron charge. For the sake of simplicity, we assume that the hopping integral is a constant parameter, independent on the external fields. 

In Hamiltonian (\ref{eq1}), $c_{\gamma,\sigma}^+$ and $c_{\gamma,\sigma}$ are the electron creation and annihilation operators, respectively, and $\sigma$ denotes the spin state. The on-site occupation number operators for given spin, $n_{\gamma,\sigma}$, are expressed by $n_{\gamma,\sigma}=c_{\gamma,\sigma}^+c_{\gamma,\sigma}$. The $z$-component of the electron spin on given atom, $S_{\gamma}^z$, is then defined as  $S_{\gamma}^z=\left(n_{\gamma,\uparrow}-n_{\gamma,\downarrow}\right)/2$. In turn, the total occupation number operators $n_{\gamma}$ for site $\gamma = a,b$, are defined as a sum of occupation operators for given spin, $n_{\gamma}=n_{\gamma,\uparrow}+n_{\gamma,\downarrow}$.

Beause of treating the pair-cluster as an open electron system within the formalism of grand canonical ensemble, the Hamiltonian should be extended by the chemical potential term, i.e., $\mathcal{H}_{a,b}-\mu \left(n_a+n_b\right)$ is considered, where $\mu$ is the chemical potential. The exact analytical diagonalization of the extended Hamiltonian has been performed in Ref. \cite{Balcerzak2017}. As a result, not only the statistical, but also thermodynamic properties can be calculated exactly. In particular, the grand thermodynamic potential $\Omega_{a,b}$ has been obtained in the form:
\begin {equation}
\Omega_{a,b}=-k_{\rm B}T \ln \mathcal{Z}_{a,b}=-k_{\rm B}T \ln \{ {\rm Tr}_{a,b} \,\exp \lbrack -\beta \left(\mathcal{H}_{a,b}-\mu\left(n_a+n_b\right)\right)\rbrack \},
\label{eq2}
\end {equation}
where $\mathcal{Z}_{a,b}$ is the grand partition function.

The self-consistent calculations of the chemical potential have also been performed in Ref. \cite{Balcerzak2017}. It is worth mentioning here that $\mu$ can be found from the relationship 
\begin {equation}
-\left(\frac{\partial \Omega}{\partial \mu}\right)_{T,H,E}=\left(\left<n_a\right>+\left<n_b\right>\right)=2x
\label{eq3}
\end {equation}
where $\left<n_a\right>$ and $\left<n_b\right>$ are the thermodynamic mean values of the total occupation number operators for $\gamma =a,b$ sites, respectively. The parameter $x$, where $0\le x\le 2$ for open system in equilibrium, denotes the mean number of electrons per lattice site, i.e., the electron concentration. The partial derivative in Eq.(\ref{eq3}) is performed at constant temperature $T$ and external fields $H$ and $E$. The statistical averages of the on-site occupation number operators in Eq. (\ref{eq3}) are independently calculated from the formula:
\begin {equation}
\left<n_{\gamma} \right>={\rm Tr}_{a,b} \left[ \left(n_{\gamma,\uparrow}+n_{\gamma,\downarrow}\right)\; \hat \rho_{a,b} \right],
\label{eq4}
\end {equation}
where $\hat \rho_{a,b}$ is the statistical operator for the grand canonical ensemble given by:
\begin {equation}
\hat \rho_{a,b}=\frac{1}{\mathcal{Z}_{a,b}}\, \exp \lbrack -\beta \left(\mathcal{H}_{a,b}-\mu\left(n_a+n_b\right)\right)\rbrack.
\label{eq5}
\end {equation}
By the same token, the on-site magnetization, $m_{\gamma}$, can be calculated as the statistical average of $z$-component of the spins, namely $m_{\gamma}=\left<S_{\gamma}^z\right>$, where
\begin {equation}
\left<S_{\gamma}^z \right>={\rm Tr}_{a,b} \left[ \frac{1}{2}\left(n_{\gamma,\uparrow}-n_{\gamma,\downarrow}\right)\; \hat \rho_{a,b} \right].
\label{eq6}
\end {equation}
Having calculated these averages, the mean magnetic polarization per one atom, $m$, is defined as:
\begin {equation}
m=\frac{1}{2}\left(m_{a}+m_{b}\right),
\label{eq7}
\end {equation}
from which the magnetic susceptibility $\chi_{H}$ can be directly obtained:
\begin {equation}
\chi_{H}=\left(\frac{\partial m}{\partial H}\right)_{T,E}.
\label{eq8}
\end {equation}
On the other hand, the electric field $E$ induces the dipolar electric moment on the Hubbard dimer. The absolute value of the electric polarization, $P$, which is proportional to the mean charge displacement and interatomic distance $d$, can be found from the formula:
\begin {equation}
P=d \;|e|\;|\left<n_a\right>-x|.
\label{eq9}
\end {equation}
On this basis, the electric response function, i.e., electric susceptibility 
$\chi_{E}$, can be found:
\begin {equation}
\chi_{E}=\left(\frac{\partial P}{\partial E}\right)_{T,H}.
\label{eq10}
\end {equation}
The above formalism will be employed as a basis for the numerical calculations  discussed in the next Section \ref{num}.\\

\section{\label{num}Numerical results and discussion}

\begin{figure}[h]
\begin{center}
\includegraphics[width=0.9\textwidth]{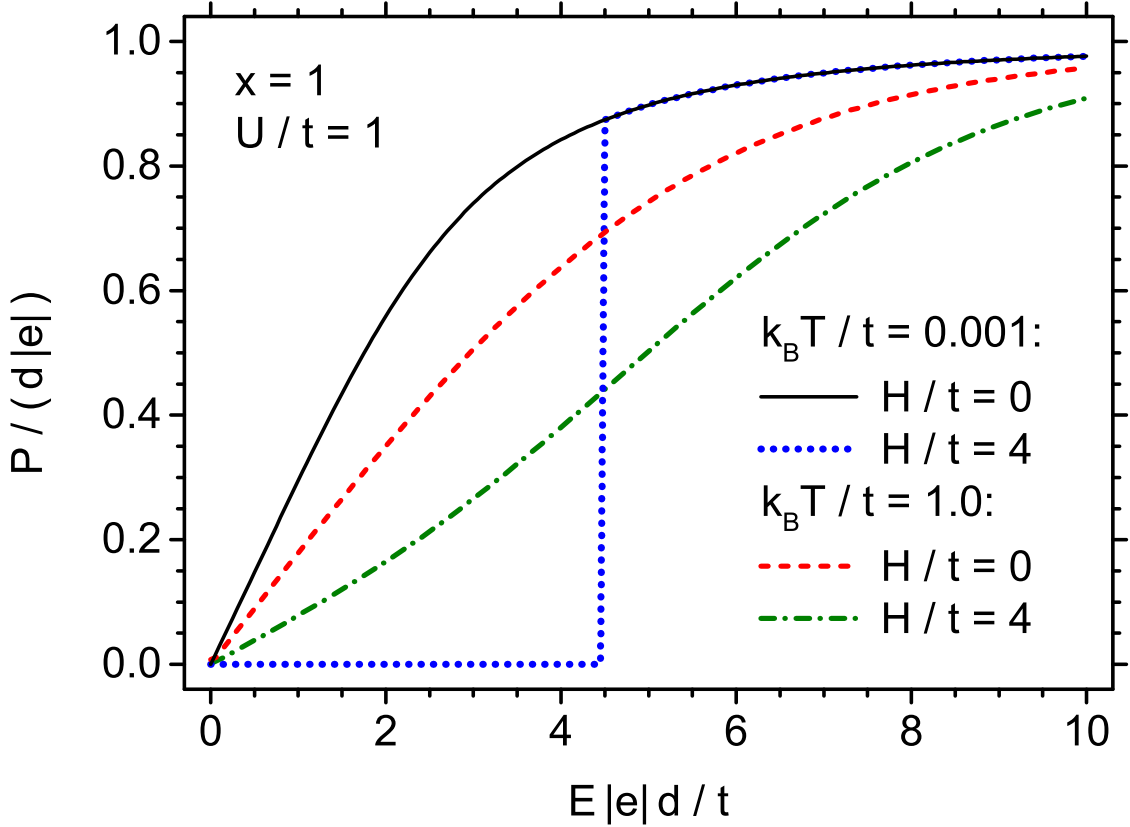}
\vspace{2mm}
\caption{\label{Fig1} The electric polarization, $P$, plotted in dimensionless units $P/\left( d |e|\right)$ as a function of the potential difference $E|e|d/t$, for $U/t=1$ and $x=1$. Different curves correspond to various dimensionless temperatures $k_{\rm B}T/t$ and magnetic fields $H/t$.} 
\end{center}
\end{figure}

\begin{figure}[h]
\begin{center}
\includegraphics[width=0.9\textwidth]{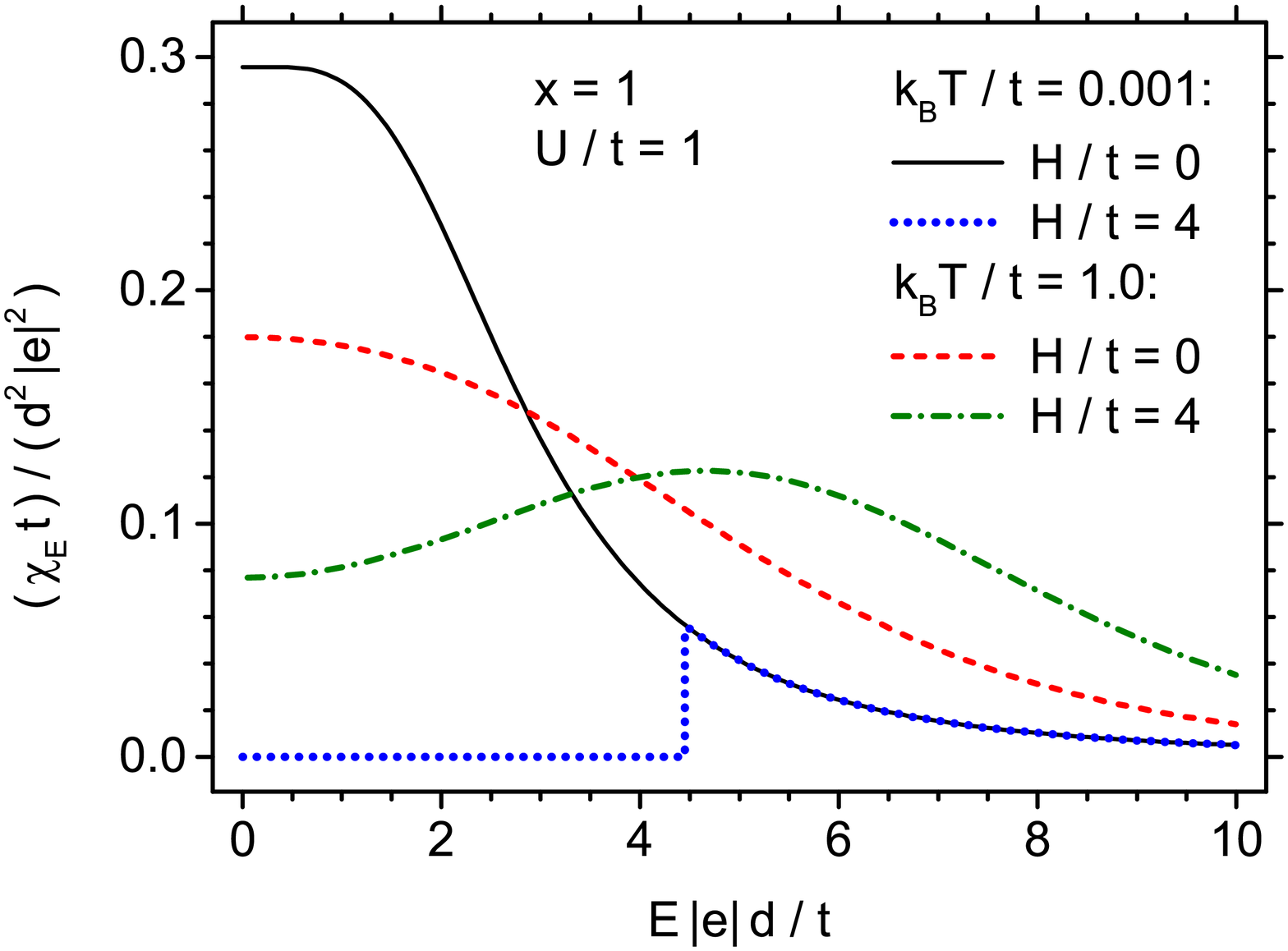}
\vspace{2mm}
\caption{\label{Fig2} The electric susceptibility, $\chi_E$, plotted  in dimensionless units $\left(t\chi_E\right)/\left( d^2 |e|^2\right)$  as a function of the potential difference $E|e|d/t$, for $U/t=1$ and $x=1$. Different curves correspond to various dimensionless temperatures $k_{\rm B}T/t$ and magnetic fields $H/t$.}
\end{center}
\end{figure}

\begin{figure}[h]
\begin{center}
\includegraphics[width=0.9\textwidth]{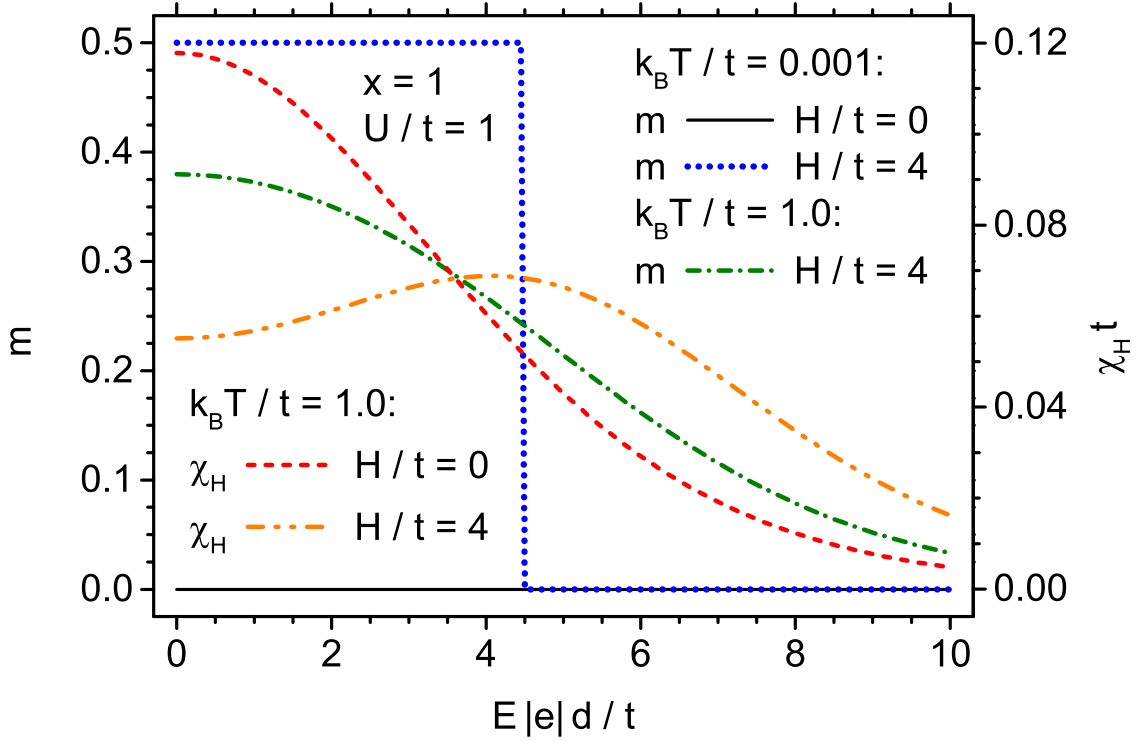}
\vspace{2mm}
\caption{\label{Fig3} The magnetic polarization (dimensionless magnetization) per atom, $m$, and magnetic susceptibility, $\chi_H$, plotted in dimensionless units $t\chi_H$ vs. potential difference $E|e|d/t$, for $U/t=1$ and $x=1$. Different curves correspond to various dimensionless temperatures $k_{\rm B}T/t$ and magnetic fields $H/t$.}
\end{center}
\end{figure}

\begin{figure}[h]
\begin{center}
\includegraphics[width=0.9\textwidth]{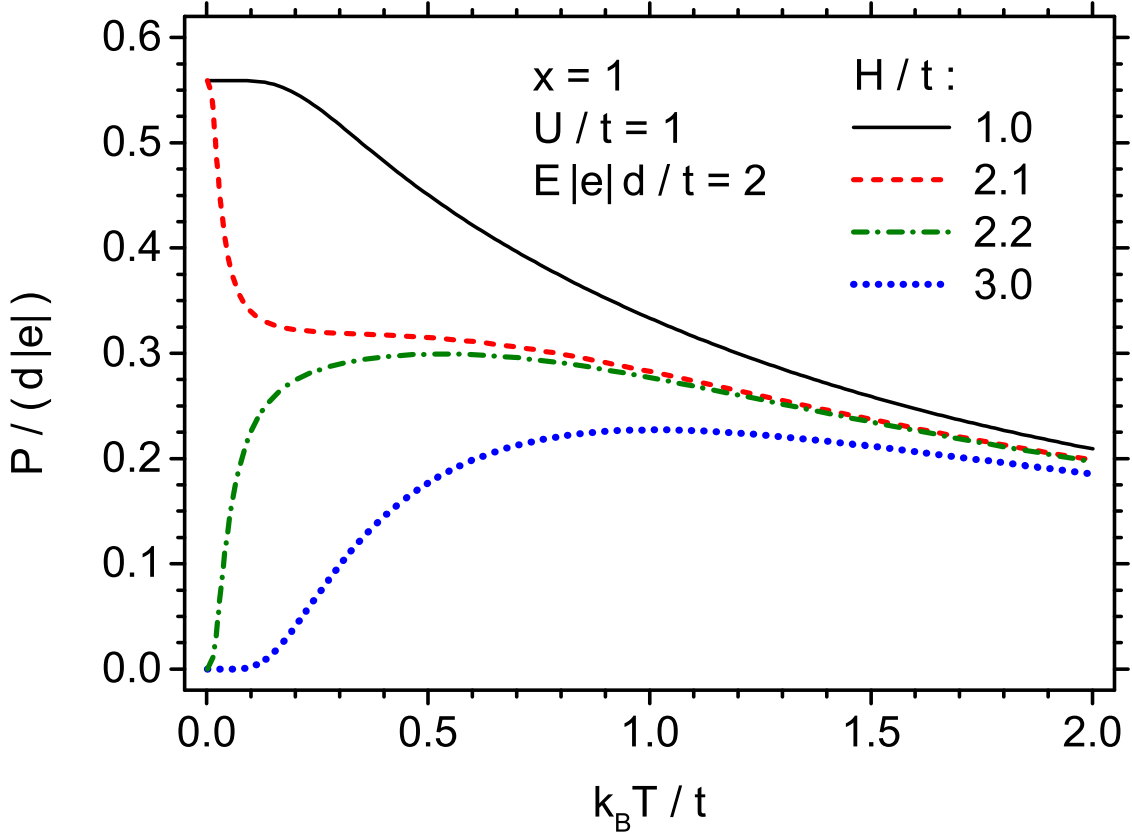}
\vspace{2mm}
\caption{\label{Fig4} The electric polarization, $P$, plotted  in dimensionless units $P/\left( d |e|\right)$ as a function of the dimensionless temperature $k_{\rm B}T/t$, for $U/t=1$, $E|e|d/t=2$ and $x=1$. Different curves correspond to various external magnetic fields $H/t$.}
\end{center}
\end{figure}

\begin{figure}[h]
\begin{center}
\includegraphics[width=0.9\textwidth]{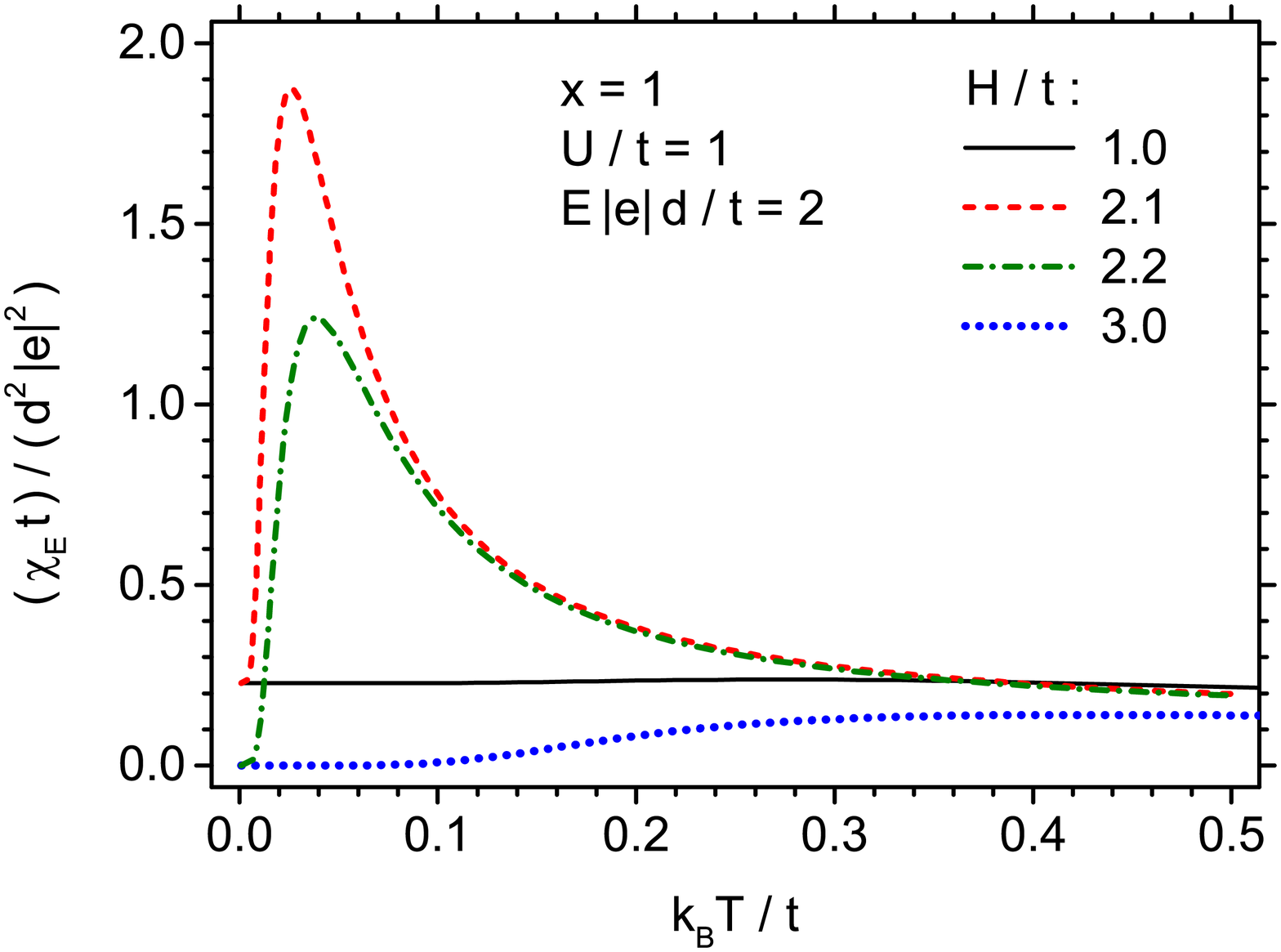}
\vspace{2mm}
\caption{\label{Fig5} The electric susceptibility, $\chi_E$, plotted  in dimensionless units $\left(t\chi_E\right)/\left( d^2 |e|^2\right)$  as a function of the dimensionless temperature $k_{\rm B}T/t$, for $U/t=1$, $E|e|d/t=2$ and $x=1$. Different curves correspond to various external magnetic fields $H/t$.}

\end{center}
\end{figure}

\begin{figure}[h]
\begin{center}
\includegraphics[width=0.9\textwidth]{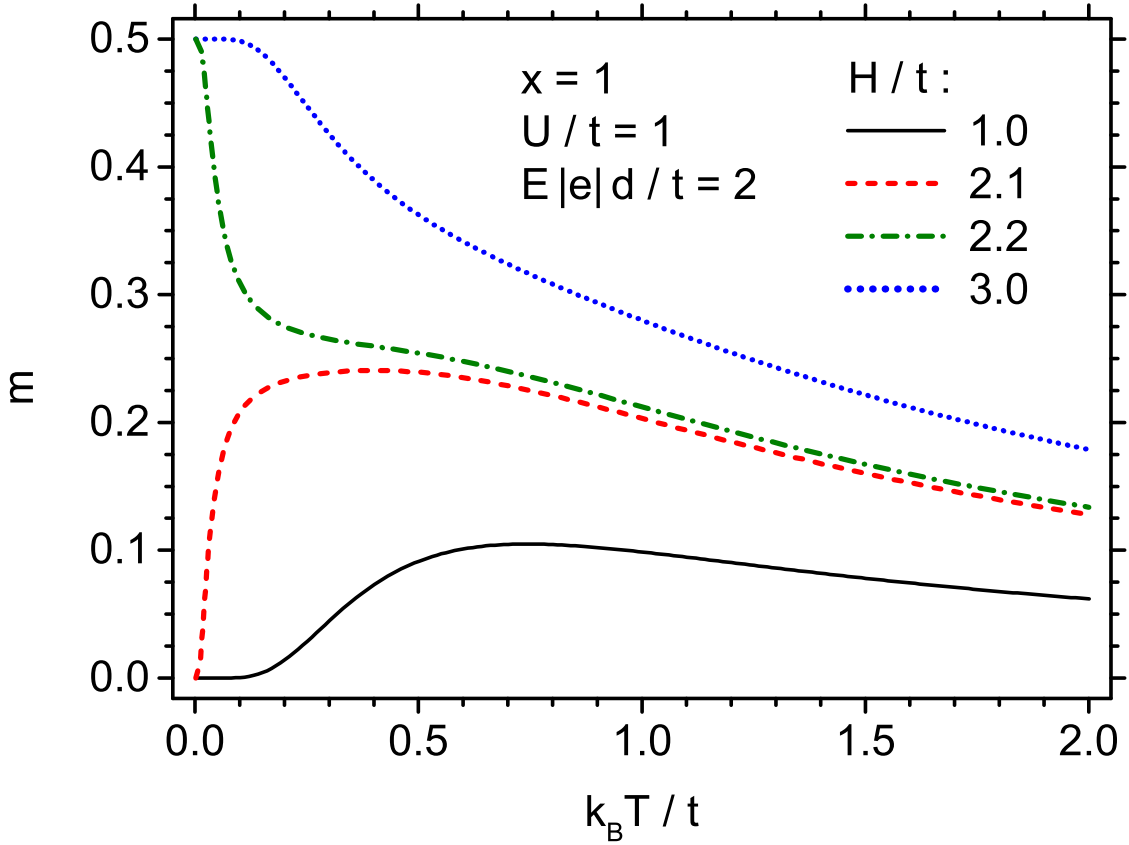}
\vspace{2mm}
\caption{\label{Fig6} The dimensionless magnetization per atom, $m$, as a function of the dimensionless temperature $k_{\rm B}T/t$, for $U/t=1$, $E|e|d/t=2$ and $x=1$. Different curves correspond to various external magnetic fields $H/t$.}
\end{center}
\end{figure}

\begin{figure}[h]
\begin{center}
\includegraphics[width=0.9\textwidth]{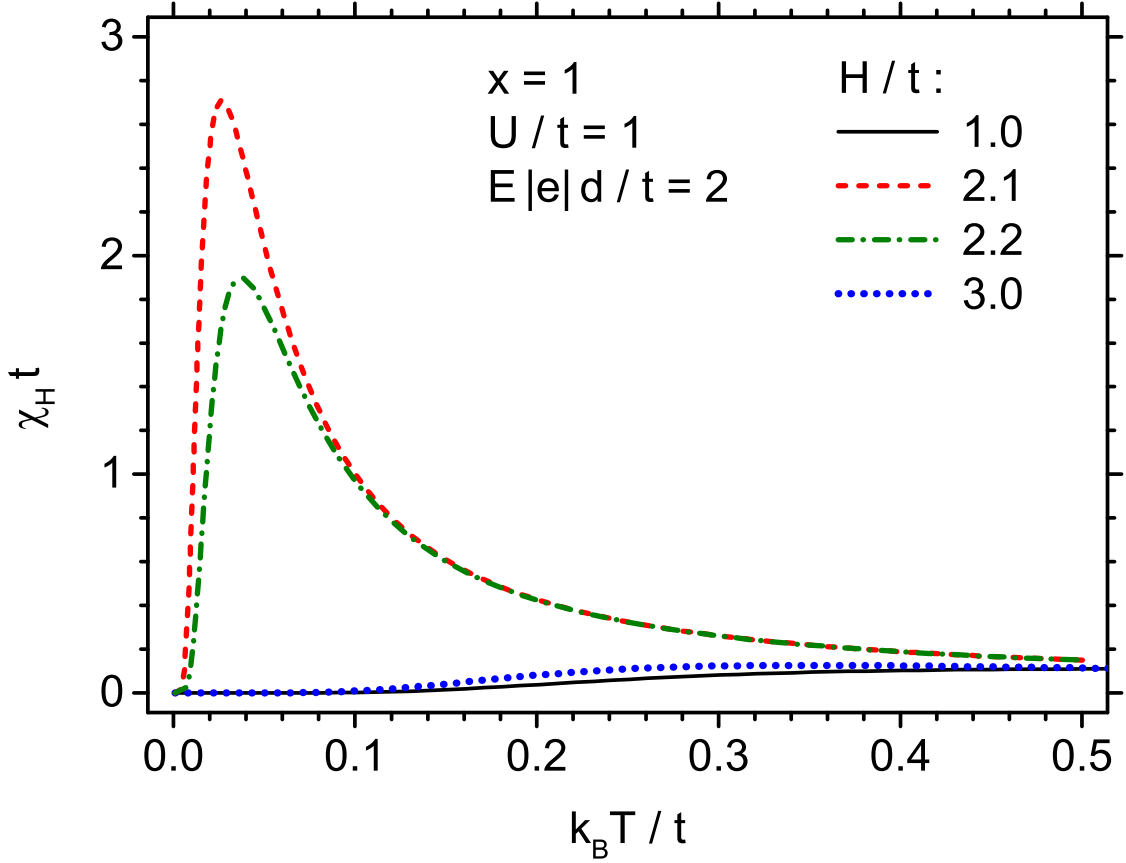}
\vspace{2mm}
\caption{\label{Fig7} The dimensionless magnetic susceptibility, $t\chi_H$, as a function of the dimensionless temperature $k_{\rm B}T/t$, for $U/t=1$, $E|e|d/t=2$ and $x=1$. Different curves correspond to various external magnetic fields $H/t$.}
\end{center}
\end{figure}

\begin{figure}[h]
\begin{center}
\includegraphics[width=0.9\textwidth]{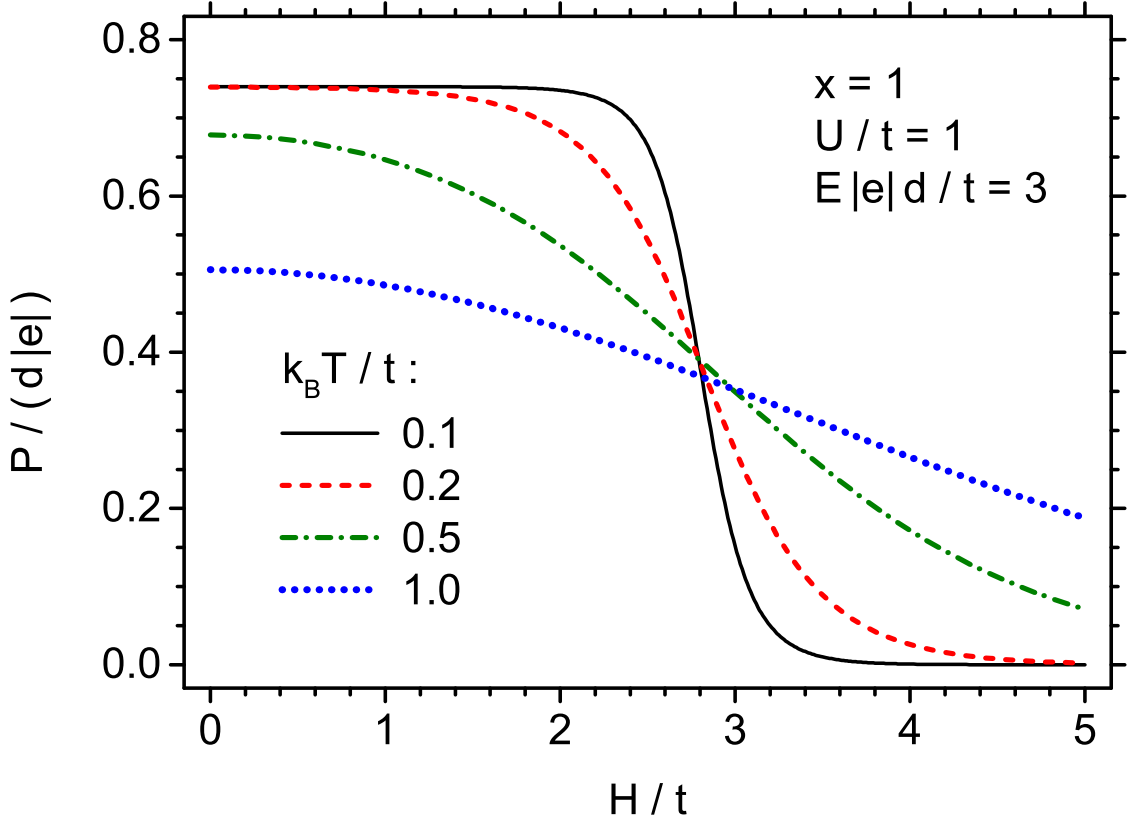}
\vspace{2mm}
\caption{\label{Fig8} The dimensionless electric polarization, $P/\left( d |e|\right)$, vs. magnetic field $H/t$, for $U/t=1$, $E|e|d/t=3$ and $x=1$. Different curves correspond to several selected temperatures $k_{\rm B}T/t$.}
\end{center}
\end{figure}

\begin{figure}[h]
\begin{center}
\includegraphics[width=0.9\textwidth]{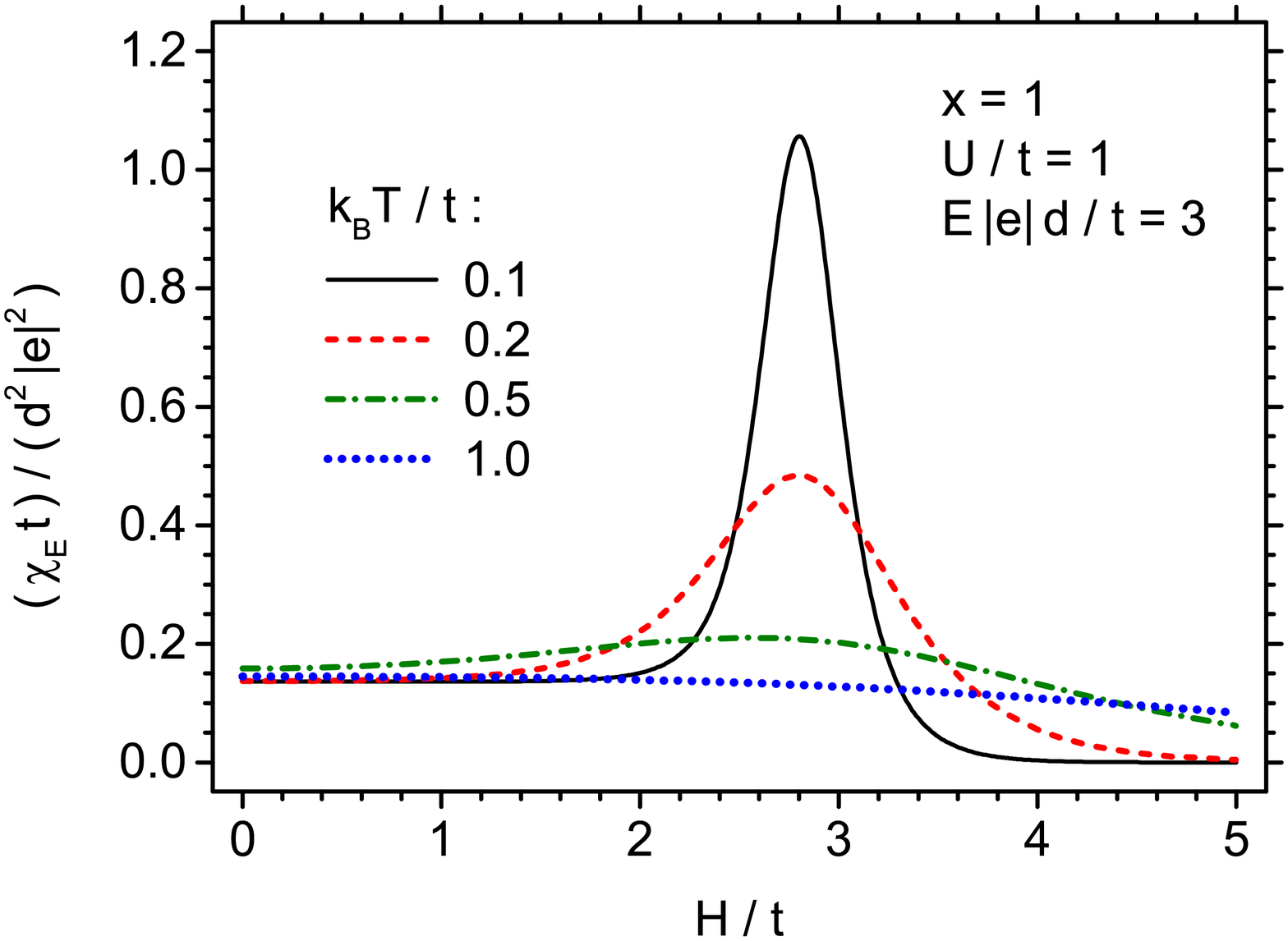}
\vspace{2mm}
\caption{\label{Fig9} The dimensionless electric susceptibility, $\left(t\chi_E\right)/\left( d^2 |e|^2\right)$, vs. magnetic field $H/t$, for $U/t=1$, $E|e|d/t=3$ and $x=1$. Different curves correspond to several selected temperatures $k_{\rm B}T/t$.}
\end{center}
\end{figure}

\begin{figure}[h]
\begin{center}
\includegraphics[width=0.9\textwidth]{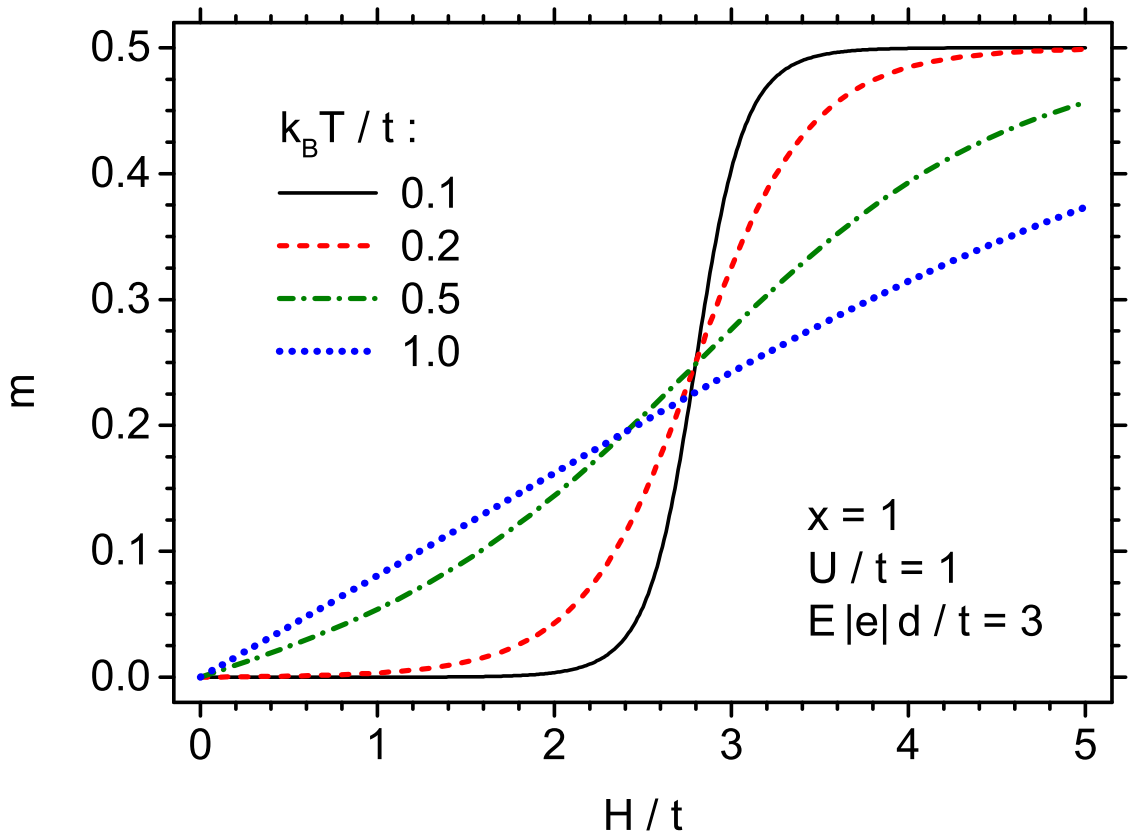}
\vspace{2mm}
\caption{\label{Fig10} The dimensionless magnetization per atom, $m$, vs. magnetic field $H/t$, for $U/t=1$, $E|e|d/t=3$ and $x=1$. Different curves correspond to several selected temperatures $k_{\rm B}T/t$.}
\end{center}
\end{figure}

\begin{figure}[h]
\begin{center}
\includegraphics[width=0.9\textwidth]{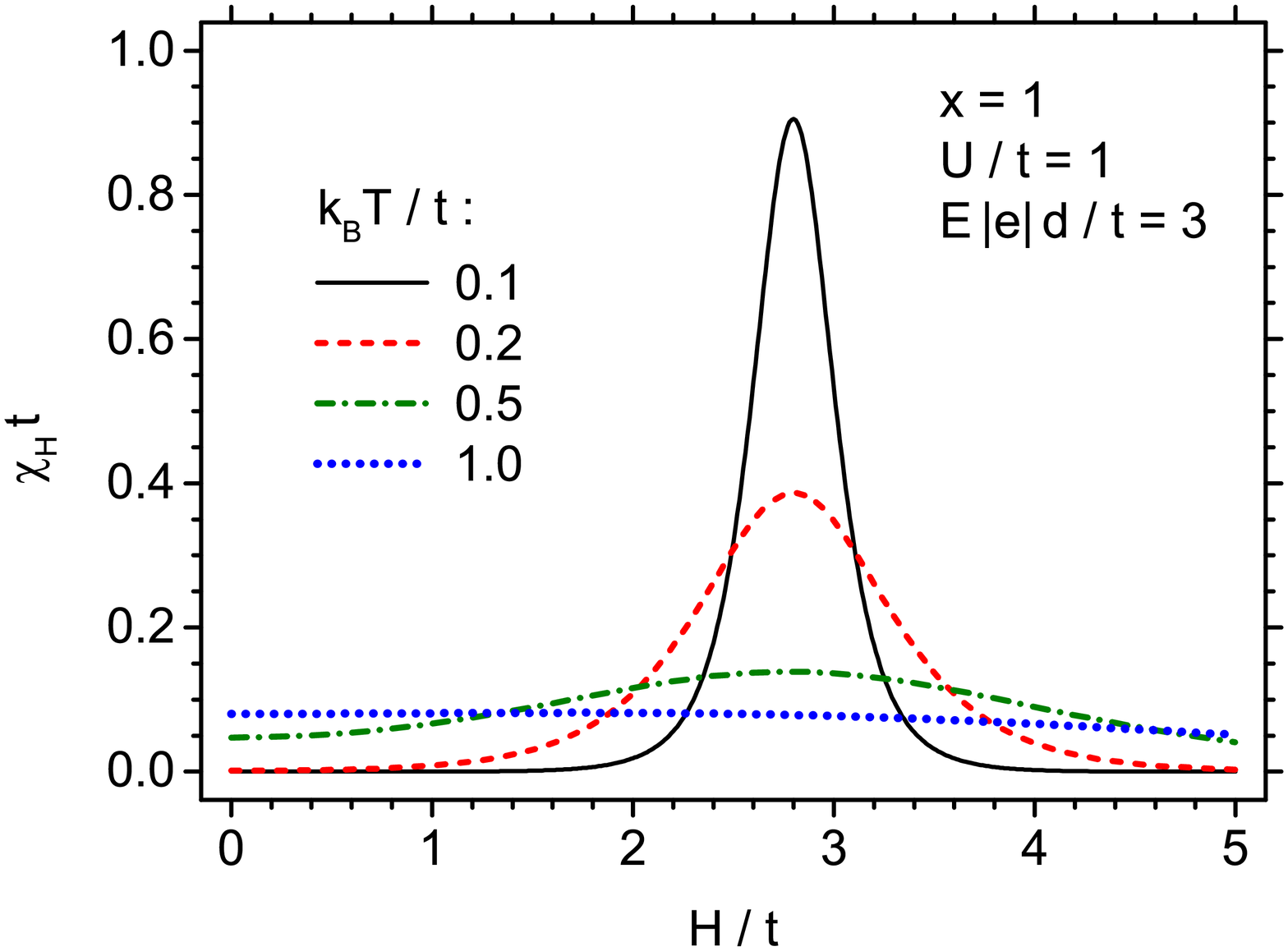}
\vspace{2mm}
\caption{\label{Fig11} The dimensionless magnetic susceptibility, $t\chi_H$, vs. magnetic field $H/t$, for $U/t=1$, $E|e|d/t=3$ and $x=1$. Different curves correspond to several selected temperatures $k_{\rm B}T/t$.}
\end{center}
\end{figure}

\begin{figure}[h]
\begin{center}
\includegraphics[width=0.9\textwidth]{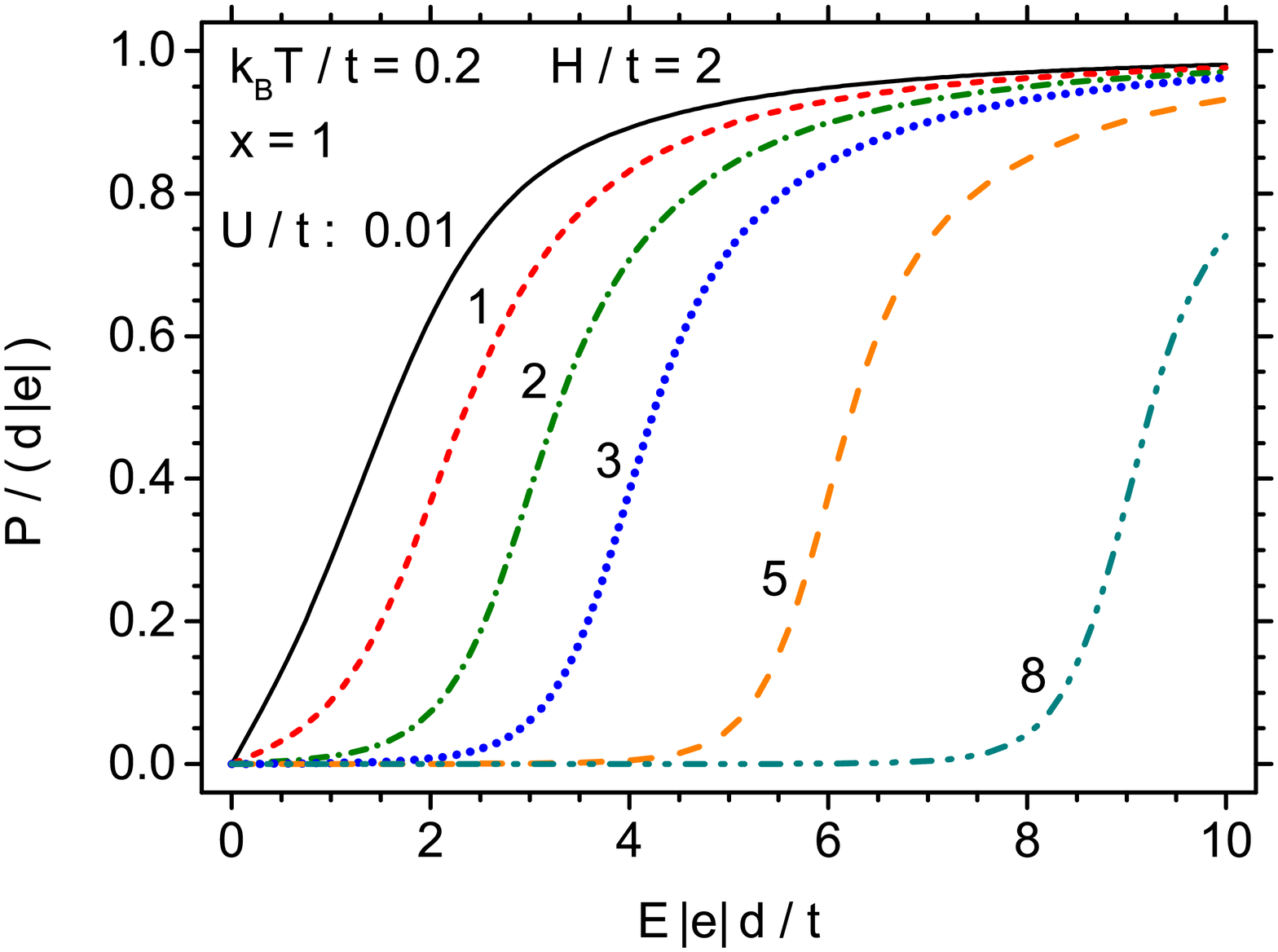}
\vspace{2mm}
\caption{\label{Fig12} The dimensionless electric polarization, $P/\left( d |e|\right)$, as a function of the potential difference $E|e|d/t$, for $H/t=2$, $k_{\rm B}T/t=0.2$ and $x=1$. Different curves correspond to various Coulomb repulsion $U/t$-parameter.}
\end{center}
\end{figure}

\begin{figure}[h]
\begin{center}
\includegraphics[width=0.9\textwidth]{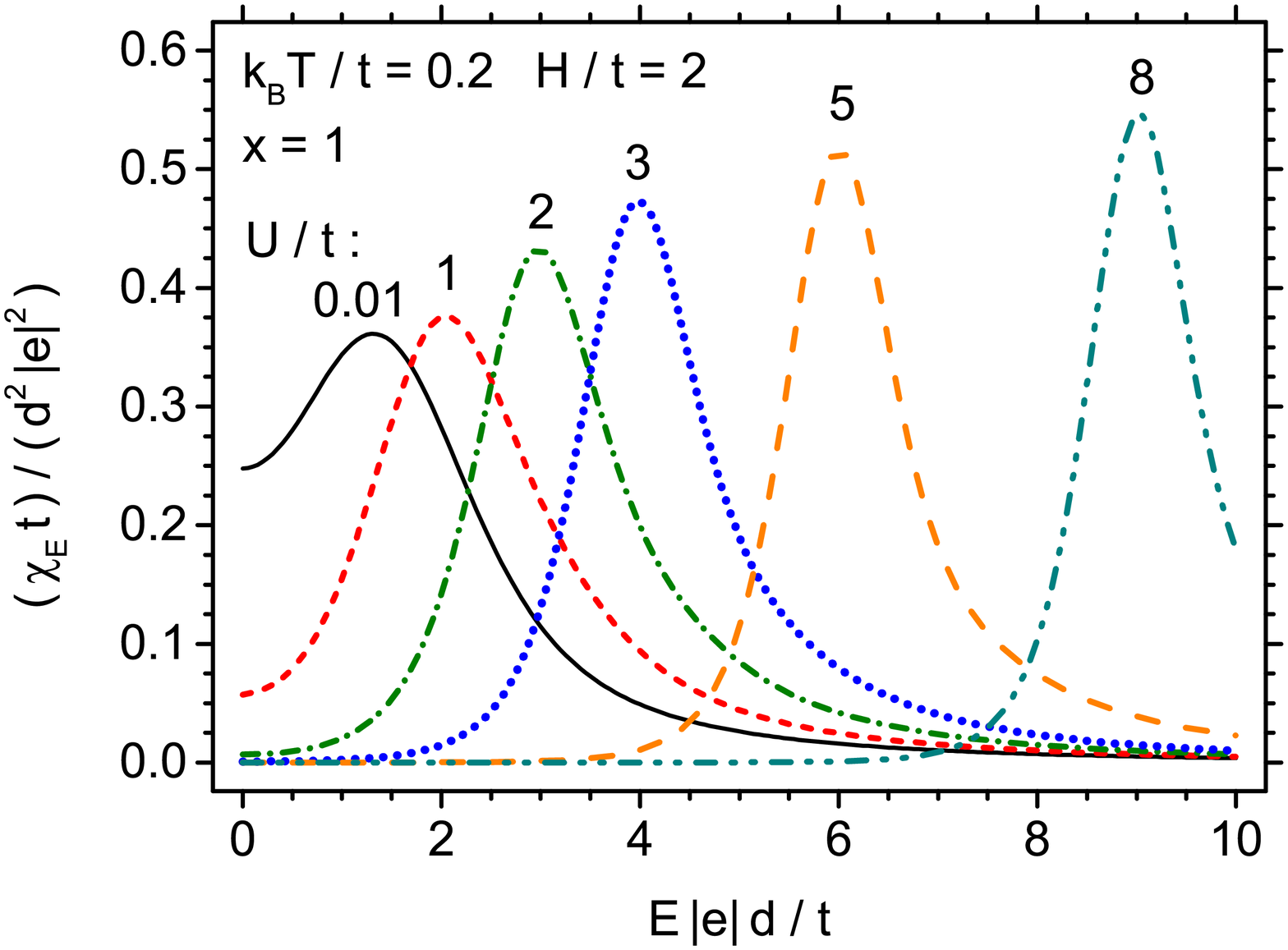}
\vspace{2mm}
\caption{\label{Fig13} The dimensionless electric susceptibility, $\left(t\chi_E\right)/\left( d^2 |e|^2\right)$, as a function of the potential difference $E|e|d/t$, for $H/t=2$, $k_{\rm B}T/t=0.2$ and $x=1$. Different curves correspond to various Coulomb repulsion $U/t$-parameters.}
\end{center}
\end{figure}

\begin{figure}[h]
\begin{center}
\includegraphics[width=0.9\textwidth]{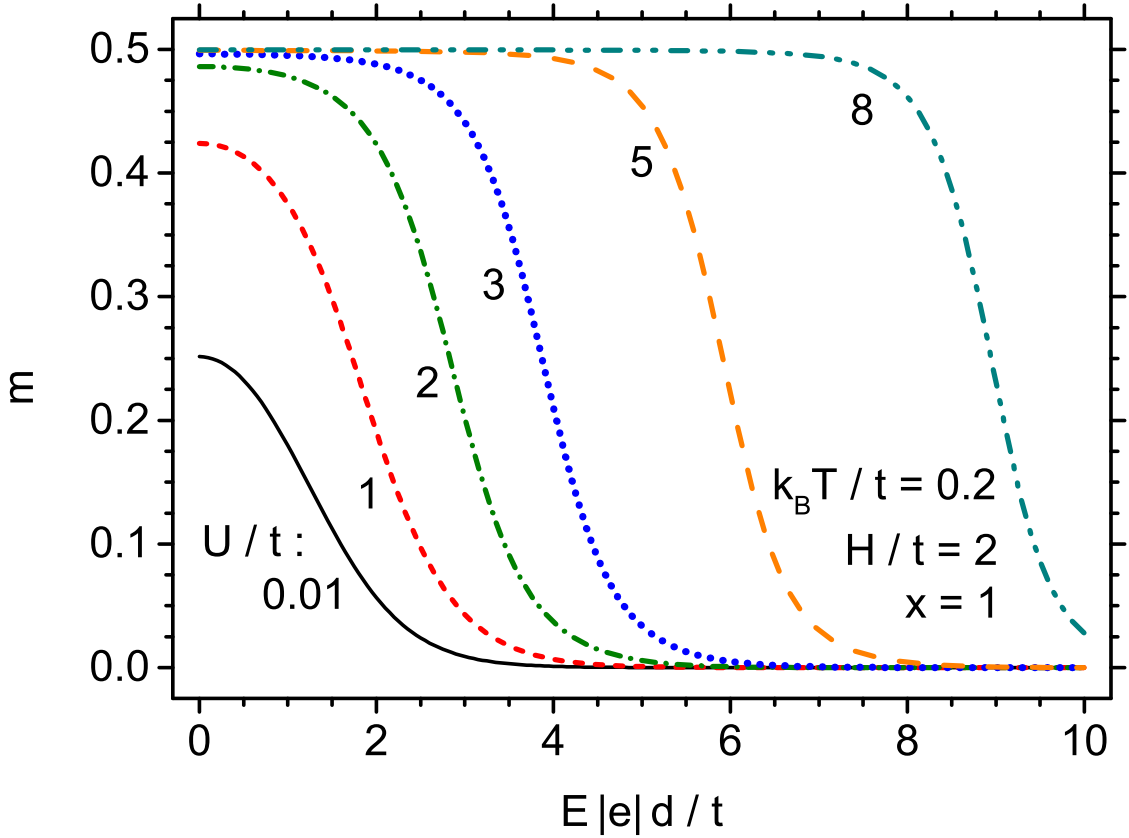}
\vspace{2mm}
\caption{\label{Fig14} The dimensionless magnetization per atom, $m$, as a function of the potential difference $E|e|d/t$, for $H/t=2$, $k_{\rm B}T/t=0.2$ and $x=1$. Different curves correspond to various Coulomb repulsion $U/t$-parameters.}
\end{center}
\end{figure}

\begin{figure}[h]
\begin{center}
\includegraphics[width=0.9\textwidth]{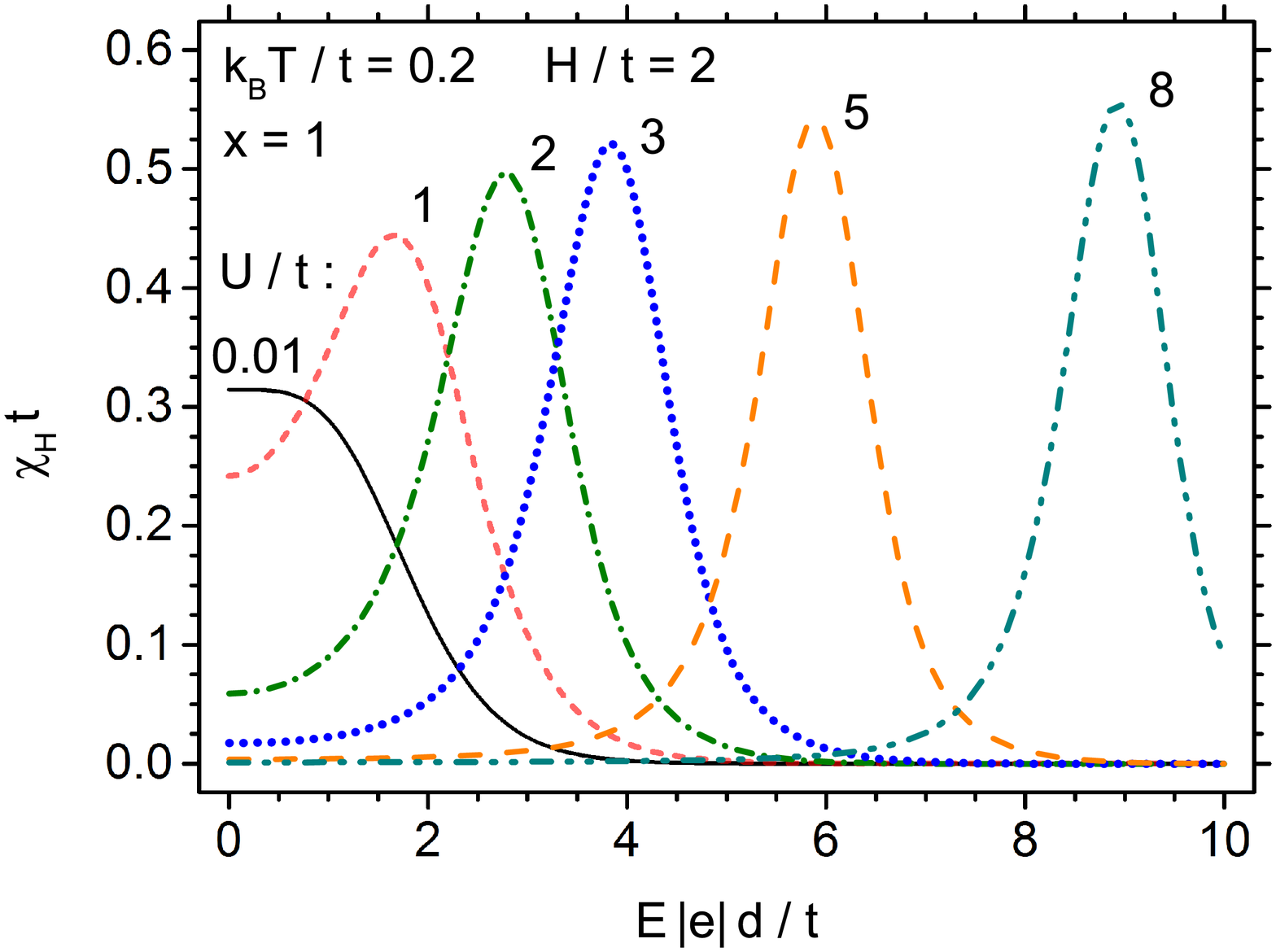}
\vspace{2mm}
\caption{\label{Fig15} The dimensionless magnetic susceptibility, $t\chi_H$, as a function of the potential difference $E|e|d/t$, for $H/t=2$, $k_{\rm B}T/t=0.2$ and $x=1$. Different curves correspond to various Coulomb repulsion $U/t$-parameters.}
\end{center}
\end{figure}

In this Section, the results of the rigorous calculations of the electric and magnetic polarization as well as the electric and magnetic susceptibilities of the Hubbard pair embedded in the external magnetic and electric fields are presented. The calculations are restricted to the most interesting case when the orbital states of Hubbard dimer are half-filled ($x=1$). 

The behaviour of both order parameters (electric polarization $P$ and magnetization $m$) as well as the behaviour of the response functions (electric and magnetic susceptibility) stems from the behaviour of the energy states of the dimer. Therefore, some microscopic insight into the physics of the studied model regarding the effect of the external fields would be valuable. For this purpose we provide the \ref{app}, which presents the energy states corresponding to half-filling of the dimer and discusses their behaviour as a function of the external electric and magnetic field. It serves as a reference point for the ground state discussion, as for $T\to 0$ only the states with two electrons per dimer are important, because the charge density fluctuations vanish.

In Fig.~\ref{Fig1} the dimensionless electric polarization $P/\left( d |e|\right)$ is plotted vs. difference of the electric potential energies $E|e|d/t$, for the electron concentration $x=1$ and Coulomb parameter $U/t=1$. Different curves correspond to two reduced temperatures: $k_{\rm B}T/t=0.001$ (i.e. system close to the ground state) and $k_{\rm B}T/t=1$, as well as two values of the magnetic fields: $H/t=0$ and $H/t=4$. Thus, the curves present the polarization process from the initial state, when the charge of two electrons is distributed equally among two atomic sites ($a$ and $b$), up to the final state, when both electrons are localized on $a$-atom (then the site $b$ is empty) and saturation of the electric polarization takes place. In particular, the curve for $k_{\rm B}T/t$=0.001 and $H/t$=4 shows a discontinuous change of electric polarization at very low temperatures when the electric field $E =  2V/\left(|e|d\right)$ exceeds some critical value. This critical value amounts to $E_c|e|d/t=4.4721$ and corresponds to the change of the ground state from a triplet one to the singlet one, as demonstrated in Fig.~\ref{Fig17} and discussed in \ref{app}. This interesting property is correlated with the magnetic polarization behaviour, as it will be explained in the discussion of Fig.~\ref{Fig3}. Comparing curves for $H/t$ = 0.0 and 4.0, plotted for the same temperature, we see that the magnetic field suppresses electric polarization. This is a kind of "spin blockade", which arises when two spins are parallel aligned in the magnetic field and they are residing on different atoms. Then, the shift of both electrons by the electric field to the same atom requires additionally the spin reversal of one electron.

The corresponding response function, electric susceptibility $\chi_E= \left(\partial P/ \partial E\right)_{T,H}$, is presented in Fig.~\ref{Fig2} in dimensionless units vs. the electric field $E|e|d/t$. The rest of parameters are the same as in Fig.~\ref{Fig1}. Speaking about the low-temperature behaviour ($k_{\rm B}T/t$=0.001), it is seen that for the magnetic field $H/t=4$ the electric susceptibility jumps from zero value up to the value corresponding to absence of the magnetic field, and this jump takes place at the same critical electric field as seen in Fig.~\ref{Fig1} and discussed above. Thus, the behaviour of electric susceptibility is closely correlated with the behaviour of electric polarization from Fig.~\ref{Fig1}. Again, comparing the curves plotted in Fig.~\ref{Fig2} for $H/t$=0.0 and 4.0, for the same temperature $k_{\rm B}T/t=1$, one can note that the external magnetic field enforces some smooth maximum of the electric susceptibility. This maximum corresponds to the point of inflection of the curve for the same parameters ($k_{\rm B} T/t=1.0$ and $H/t$=4.0) in Fig.~\ref{Fig1}. The maximum occurring on the curve for $k_{\rm B}T/t$=1.0 and $H/t$=4.0 can also be treated as a high-temperature vestige of the low-temperature sharp maximum at the discontinuous transition seen on curve for the same $H$ and $k_{\rm B}T/t$=0.001. For very large electric fields $E\propto V$, when the saturation of electric polarization takes place, the electric susceptibility approaches the zero value.

In Fig.~\ref{Fig3} the magnetic polarization (i.e., averge magnetization) per one atom, $m=\left(\left<S_a^z\right>+\left<S_b^z\right>\right)/2$, as well as the magnetic susceptibility, $\chi_H= \left(\partial m/ \partial H\right)_{T,E}$, are simultaneously presented as the functions of $E|e|d/t$. The parameters $U/t=1$ and $x=1$ are the same as in Figs.~\ref{Fig1} and \ref{Fig2}. The curve plotted in this figure for $k_{\rm B}T/t=0.001$ and $H/t=4$ shows the first order magnetic transition caused by the electric field (and corresponds to the curves for the same parameters in Figs.~\ref{Fig1} and ~\ref{Fig2} also exhibiting the discontinuous behaviour). For low values of $E|e|d/t$, the state of the system is a triplet one, as discussed in the \ref{app}, with the spins of both electrons aligned along the magnetic field $H$ if its magnitude exceeds the critical value (see discussion of this effect in our previous paper \cite{Balcerzak2018}). The electrons are then localized on $a$ and $b$ atoms and the electric polarization is zero (see Fig.~\ref{Fig1}). When the critical electric field is reached in low temperatures, both electrons are forced to localize on the same atom, which results in rapid increase of polarization (see Fig.~\ref{Fig1}), but at the same time the spin of one electron must be reversed and the magnetization discontinuously drops to zero. On the other hand, for curve plotted near the ground state ($k_{\rm B}T/t$=0.001) and for $H/t$=0, the lack of magnetic polarization is observed in the whole range of electric field $E$. Since in this case the magnetic field $H$ is absent, thus the state of the system is a singlet and the electrons with opposite spins can freely occupy both atoms with the same probability. Increasing the temperature will not change this nonmagnetic state. For instance, the magnetic susceptibility, illustrated in the case of $k_{\rm B}T/t=1$ and $H/t=0$ shows typical paramagnetic behaviour. However, when the strong magnetic field $H/t=4$ is applied at high temperatures, for instance, at $k_{\rm B}T/t=1$, the behaviours of magnetization and magnetic susceptibility are very different from those at the ground state. Namely, magnetic polarization decreases as a function of the field $E$, whereas magnetic susceptibility reveals a broad maximum, whose position is correlated with the magnetic transition observed in the ground state. We note that diminishing of magnetic polarization vs. electric field, as seen in curve for $k_{\rm B}T/t$=1.0 and $H/t$=4.0, is also correlated with the increase of electric polarization illustrated in Fig.~\ref{Fig1}. The continuous (and smooth) curves, observed for high temperatures, are connected with the statistical averaging of occupancy of all electronic states. The electric field-dependence of the magnetization demonstrated in Fig.~\ref{Fig3} is a clear example of magnetoelectric effect, allowing the control of magnetization by electric means. 

Figs.~\ref{Fig4}, \ref{Fig5}, \ref{Fig6} and \ref{Fig7} present the electric polarization, electric susceptibility, magnetization and magnetic susceptibility, respectively. The plots show the temperature dependence of the mentioned quantities and are prepared for the same remaining parameters: $x=1$, $U/t=1$ and $E|e|d/t=2$, as well as for the magnetic fields: $H/t=1$, $H/t=2.1$, $H/t=2.2$ and $H/t=3$. For the above set of parameters the critical magnetic field has been found as $H_c/t=2.1411$ (see the discussion in the \ref{app} concenrning the critical field for the transition between the singlet and triplet state). Thus, two of the curves fall into the range below the critical field (i.e. correspond to the singlet at the ground state) and two other are shown for $H>H_c$ (and correspond to a triplet at the ground state). It can be mentioned here that, as discussed in the paper \cite{Balcerzak2018}, the critical magnetic field $H_c$ is the field above which the system presents ferromagnetic ordering in the ground state, whereas for $H<H_c$ the paramagnetic (singlet) ground state exists. 

Analysing the Fig.~\ref{Fig4}, which shows the dimensionless polarization $P/\left( d |e|\right)$, we see that below $H_c$ (curves for $H/t$ =1.0 and 2.1) the electric polarization in the field $E$ reaches a maximum for $T=0$, whereas above the critical magnetic field $H_c$ (curves plotted for $H/t$ = 2.2 and 3.0) it takes the zero value at the ground state. The jump of electric polarization, when $H_c$ is crossed, is evidently connected with the change of spin state and jump of the magnetization $m$, as seen in Fig.~\ref{Fig6}. When temperature increases, all the curves in Fig.~\ref{Fig4} converge and the influence of magnetic field becomes negligible. The same can be said about the magnetization curves in Fig.~\ref{Fig6}, where for $T\to \infty$ the magnetic polarization tends to zero, however, the convergence in that case is much slower. The fact that both electric and magnetic polarizations are very slowly tending to zero when $T$ is large, will have consequences to the existence of non-vanishing corresponding susceptibilities.

The behaviour of dimensionless electric susceptibility, $\chi_E$, vs. temperature, is illustrated in Fig.~\ref{Fig5}. First of all, in the ground state the electric susceptibility is different in two ranges: $H<H_c$ and $H>H_c$. When temperature slightly increases, narrow maxima appear for the curves plotted in the vicinity of the critical magnetic field (i.e. for $H/t$=2.1 and 2.2). During further increase in temperature all the curves become mutually convergent, independently on the magnetic field strength. It can be noted that the jump of the curves in Figs.~\ref{Fig4} and \ref{Fig5} from non-zero value (for $H<H_{c}$) to zero value (for $H>H_{c}$) when $T \to 0$ means that both the electric polarization and the electric susceptibility are vanishing in the ferromagnetic (triplet) ground state. This reflects the fact that two electrons with the same spin cannot be localized at the same atom. At the ground state there is no contribution from other states than these with two electrons per dimer, as charge density fluctuations vanish.

Regarding Fig.~\ref{Fig6}, in which the mean magnetic polarization $m=\left(\left<S_a^z\right>+\left<S_b^z\right>\right)/2$ is shown, it is worth noticing that the curves plotted for $H<H_{c}$ (i.e. for $H/t$=1.0 and 2.1) show an anomalous behaviour vs. temperature in a form of a broad maximum. It can also be noted that Fig.~\ref{Fig6} is qualitatively similar to Fig.~\ref{Fig4}, however, the curves with the same $H$-parameters are arranged in the inverse order. It should be mentioned that similar anomalous behaviour of the magnetization vs. temperature has been found in Ref. \cite{Ricardo-Chavez2001} (Fig.2 in \cite{Ricardo-Chavez2001} for $U/t$=16) in case of 6-site cluster, but without external fields.

The magnetic susceptibility, $\chi_H$, for the system embedded in the electric field $E$, is plotted vs. temperature in Fig.~\ref{Fig7}. Also in this case the pronounced maxima for the curves near the critical magnetic field $H_c$ ($H/t$ = 2.1 and 2.2) can be observed in the low-temperature region. The positions of these maxima are correlated with the most rapid changes of the magnetization presented in Fig.~\ref{Fig6} which manifest themselves for the magnetic field close to the critical field for the transition from singlet to triplet state. Similarly to Fig.~\ref{Fig5}, the magnetic susceptibility curves presented here show also the mutual convergence when temperature increases. Moreover, it should be noted that in the ground state the magnetic susceptibility always goes to zero, irrespective of the magnetic field.

In Figs.~\ref{Fig8}-\ref{Fig11} the same quantities are presented as in Figs.~\ref{Fig4}-\ref{Fig7}, but now the dependencies are shown vs. external magnetic field $H/t$, whereas the electric field is constant. The remaining parameters are: $x=1$, $U/t=1$ and $E|e|d/t=3$. For this set of parameters the critical magnetic field amounts to $H_c/t=2.7986$ (for discussion see \ref{app}). Different curves in Figs.~\ref{Fig8}-\ref{Fig11} correspond to four selected temperatures: $k_{\rm B}T/t$=0.1, 0.2, 0.5, and 1.0.

In Fig.~\ref{Fig8} the electric polarization, $P/\left( d |e|\right)$, is shown as a function of the magnetic field, $H/t$, for the temperatures specified above. The electric polarization diminishes with the increase of the magnetic field, and the most rapid changes occur in the vicinity of the critical field $H_c$. This fact confirms our previous observation that the change of magnetic ordering inevitably influences the electric polarization so that the system shows clear magnetoelectric properties. When the temperature increases, the curves flatten, showing a decrease of electric polarization for low magnetic fields, however, in the region of strong fields and $H>H_c$ some increase of the electric polarization with increasing temperature can be observed. 
   
The behaviour of electric susceptibility, $\chi_E$, vs. $H/t$ can be analysed on the basis of Fig.~\ref{Fig9}. For low temperatures ($k_{\rm B}T/t$=0.1 and 0.2) a strong peak appears at the critical magnetic field $H_c$. For higher temperatures the peak diminishes and becomes increasingly diffused. It can be noted that for $H=0$ the electric susceptibility takes a non-zero value and hardly depends on the temperature. This fact is connected with a strong electric polarization in this region, as shown in Fig.~\ref{Fig8}. On the other hand, for $H/t \to \infty$ the electric polarization tends to zero value, since the system approaches the magnetic saturation state, in which the electrons are spatially separated and localized on different atoms.  

The above discussion concerning electric properties can be extended by the analysis of magnetization and magnetic susceptibility. In Fig.~\ref{Fig10} the mean magnetic polarization, $m=\left(\left<S_a^z\right>+\left<S_b^z\right>\right)/2$, is shown as a function of $H/t$. An increasing $m$ as a function of $H$ describes the magnetization process at different temperatures. It is seen that the most rapid changes in magnetization occur near the critical magnetic field $H_c/t=2.7986$. Moreover, the changes are mostly evident for low temperatures ($k_{\rm B}T/t$=0.1), whereas for high temperatures ($k_{\rm B}T/t$=1.0) the dependence becomes weaker and tends to be  linear in the presented range of magnetic fields. In the range where $H<H_c$ an increase of the magnetization with temperature can be predicted, which supports the anomalous behaviour discussed previously and shown in Fig.~\ref{Fig6}. On the other hand, for $H>H_c$ the magnetic polarization tends to saturation value, $m=0.5$, although an increasing temperature makes this process slower.  

The behaviour of magnetic susceptibility, $\chi_H$, vs. field $H/t$ is shown in Fig.~\ref{Fig11} and it resembles, to some extent, the electric susceptibility from Fig.~\ref{Fig9}. Again, the most pronounced maxima occur near the critical magnetic field, $H_c$, and in the range of low temperatures. For $H=0$ the magnetic susceptibility approaches zero value when $T\to 0$. On the other hand, for $H\to \infty$, the magnetic susceptibility tends to zero at any finite temperature, since the system reaches the magnetically saturated (triplet) state.

In up to now presentation of the results we have assumed that the Coulomb repulsion parameter is equal to $U/t=1$. Now we will study how the variation of this parameter influences the discussed properties. In Figs.~\ref{Fig12}-\ref{Fig15} the same quantities as in previous figures are plotted vs. external electric field. The constant parameters are: $x=1$, $H/t=2$ and $k_{\rm B}T/t=0.2$. Different curves in these figures correspond to 6 values of the Coulomb parameter: $U/t$ = 0.01, 1, 2, 3, 5 and 8. 

Starting from the electric polarization, $P$, which is presented in Fig.~\ref{Fig12} in dimensionless units as a function of $E|e|d/t$, we see that with an increase in $U$-parameter the polarization curves become markedly shifted towards larger electric fields. However, the characteristic shape of these curves remains the same. Namely, for $E|e|d/t=0$ all the curves start from zero value, then, for $E|e|d/t>0$ they are increasing functions of the electric field, and eventually all of them tend to saturation polarization whereas $E|e|d/t\to \infty$. It is worth noticing that similar behaviour of the ground-state occupation difference vs. $E$, plotted for various $U$-parameters, has been found in Ref. \cite{Carrascal2015} (Fig.~4 in \cite{Carrascal2015}), but without magnetic field. The most rapid changes of polarization in Fig.~\ref{Fig12} occur for the electric fields $E$ corresponding to the critical field for singlet-triplet transition reached for the given field $H$ (see the discussion in the \ref{app}). It should be mentioned that the dependence of $H_c$ on the electric field for various $U$-parameters has been discussed in our previous paper \cite{Balcerzak2018}. It has been shown there that, for increasing $U$-parameter and constant $H=H_c$, the electric field corresponding to the transition is always shifted towards larger values. This phenomenon, resulting from the dependence of the eigenenergies of singlet and triplet states on the external fields, is nicely confirmed in the present figure. From Fig.~\ref{Fig12} we can conclude that the role of $U$-parameter consists in countering the electric polarization, as it can be expected from the model Hamiltonian, since this parameter prevents the charge localization on a single atomic site.

In Fig.~\ref{Fig13} the electric susceptibility, $\chi_E$, is plotted in dimensionless units for the same parameters as in Fig.~\ref{Fig12}. The pronounced peaks of the susceptibility can be seen for the electric fields corresponding to the singlet-triplet transitions. At the same time, the positions of these peaks reflect the points of inflection seen on the curves in Fig.~\ref{Fig12}. It can be noted that with increase of $U$-parameter the maxima of $\chi_E$ become higher and sharper. For the electric fields far from the maxima the electric susceptibilities tend to zero value, provided the $U$-parameter is high enough. On the other hand, for the range of small $U$ (like $U/t$= 0.01) and $E|e|d/t\to 0$, the susceptibility is non-zero, because of the dominating role of temperature, which here amounts to  $k_{\rm B}T/t=0.2$ (and the energy of the thermal fluctuations exceeds $U$-energy).

The mean magnetic polarization of the Hubbard pair, $m=\left(\left<S_a^z\right>+\left<S_b^z\right>\right)/2$, is presented in Fig.~\ref{Fig14} as a function of the electric field $E$. Similarly to electric polarization, the $U$-parameter has also pronounced influence on the $m$ vs. $E|e|d/t$ curves. For given $U$, magnetization diminishes from its maximal value at $E|e|d/t=0$ towards zero value when $E|e|d/t\to \infty$. The steepest decrease is observed for the same values of $E|e|d/t$ for which the electric polarization showed the steepest increase (in Fig.~\ref{Fig12}). When $U$-parameter is strong enough, the magnetization for $E|e|d/t\to 0$ is in saturated state. However, for small $U$, for example $U/t$=0.01, the magnetization cannot reach the saturation, because it is disordered by the thermal fluctuations. In general, an increase in $U$-parameter extends the range of $E|e|d/t$ in which the magnetic polarization remains in saturated state and only weakly depends on the electric field.

Finally, in the Fig.~\ref{Fig15} the magnetic susceptibility, $\chi_H$, is presented for the same parameters as in Figs.~\ref{Fig12}-\ref{Fig14}. For most of the curves, when $U$-parameters are strong enough, the distinct maxima are shown, whose character is similar to those seen in the electric susceptibility curves (Fig.~\ref{Fig13}). As discussed previously, these maxima can be attributed to the singlet-triplet transitions in the ground state, and far from these regions the magnetic susceptibility practically does not depend on the electric fields and eventually vanishes. The vanishing of $\chi_H$ occurs both in the magnetic saturation region (for $E|e|d/t\to 0$) and in the electric polarization saturation region (for $E|e|d/t\to \infty$). However, when $U$ parameter is small, for instance, for $U/t$ = 0.01 or 1, the magnetic saturation is not reached for $E|e|d/t\to 0$ at present temperature ($k_{\rm B}T/t=0.2$) and magnetic field ($H/t=2$). This fact results in a non-zero value of magnetic susceptibility for $E|e|d/t\to 0$. Moreover, in this region the positions of the magnetic susceptibility maxima, seen on curves plotted for $U/t$ = 0.01 or 1, are not coincident with the maxima of electric susceptibility (seen in Fig.~\ref{Fig13}). Thus, one can conclude that for the temperatures $T$ far from the ground state and small $U$-parameters, i.e., when the energy of the thermal fluctuations dominates, a maximum of the magnetic susceptibility is shifted with respect to the corresponding maximum of the electric susceptibility. It should be emphasized that at finite temperatures also the states higher in energy than the ground state contribute to the behaviour of the system and no sharp transition between singlet and triplet state is seen when the fields are varied. Therefore, the maxima of both susceptibilities need not to coincide exactly.

\section{\label{conc}Summary and conclusion}

In the paper the electric polarization and electric susceptibility, as well as the magnetic polarization and magnetic susceptibility have been studied for the Hubbard pair-cluster (dimer) embedded simultaneously in the electric and magnetic fields. The electron concentration has been selected at the half-filling level, i.e., with one electron per atom, since in this case the magnetic properties are most sensitive to the external fields \cite{Balcerzak2018}.
The analytical method developed in \cite{Balcerzak2017} was utilized, which enabled exact calculations for the model Hamiltonian. The present paper is a continuation of our previous work \cite{Balcerzak2018} which has been focused solely on the studies of the magnetic properties. In this work we extend the studies to include electric polarization, as well as the electric and magnetic response functions, i.e., susceptibilities. A special attention has been drawn to search of mutual correlations between the electric and magnetic properties (manifestations of the magnetoelectric effect) for the quantities mentioned above.

It has been found that the electric and magnetic polarizations, as well as the electric and magnetic susceptibilities, are strictly interrelated. For instance, an increase in the electric polarization is accompanied by corresponding decrease in total magnetization. This sort of behaviour can be generally traced back to the behaviour of singlet and triplet quantum states, which are the only states involved at the ground state (see \ref{app}). Namely, the nonmagnetic singlet state, in which the spin average vaue is zero at both sites, allows the charge redistribution under the action of the electric field and emergence of nonzero eletric polarization. On the contrary, for triplet state which exhibits ferromagnetic polarization, both electrons have parallel spins, so that they have to occupy strictly different sites and no redistribution is allowed in the electric field, preventing the electric polarization.  It has also been found that the discontinuous transition (traced back to the transition between singlet and triplet state) can be registered not only by the magnetic quantities, but also as the jumps of electric polarization or electric susceptibility. The phenomena mentioned above have been widely studied in various external fields $H$ and $E$, as well as vs. temperature $T$ for different Coulomb on-site repulsion parameter $U$. In particular, an anomalous behaviour of the electric and magnetic polarization has been found in some region of model parameters, showing the wide maxima of these quantities as a function of the temperature. The existence of the critical magnetic field $H_c$ enforcing the singlet-triplet transitions has been confirmed, and its dependence on the electric field $E$ and $U$-parameter is in accordance with our previous results \cite{Balcerzak2018}.  

Simultaneous application of the electric and magnetic fields on the system turned out to be fruitful, since it revealed a competing character of these fields. We hope that the present simple model, which has been solved exactly, can serve not only as a theoretical toy model, but it will also enable better understanding of the Hubbard model itself, and can elucidate competing interrelations between the electric and magnetic properties in strongly correlated systems. Moreover, demonstrating a clear magnetoelectric effect, it might show high potential for applications.

Further studies of the pair-cluster can be done within the so-called extended Hubbard model, when the Coulomb repulsion between electrons residing on different atomic sites is taken into account. In addition, the magnetic exchange interaction between  nearest-neighbour spins can be considered in the presence of an arbitrary electron concentration. Such generalizations of the model in question might significantly extend the range of observed phenomena.\\

\appendix

\section{\label{app}Eigenenergies of the quantum states for the dimer at half-filling}

\begin{figure}[h]
\begin{center}
\includegraphics[width=0.9\textwidth]{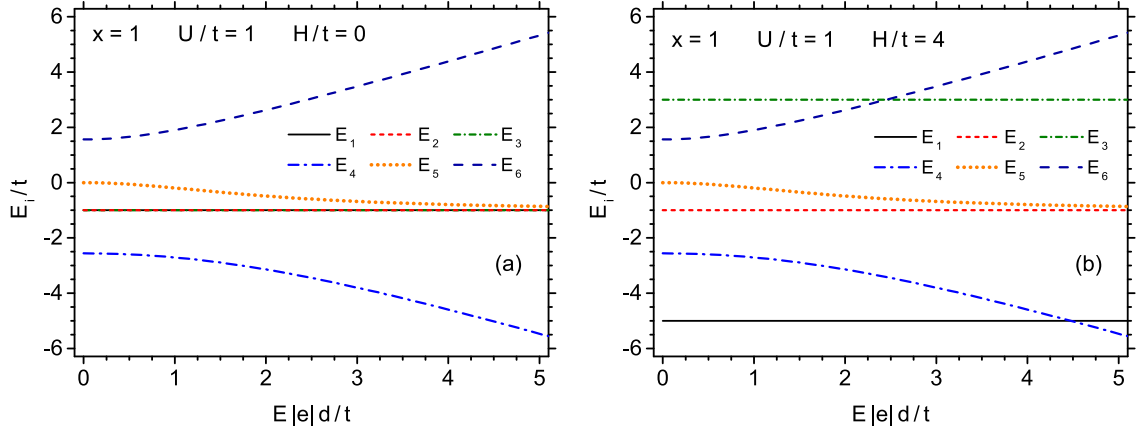}
\vspace{2mm}
\caption{\label{Fig16} The eigenenergies of the 6 quantum states corresponding to two electrons per dimer as a function of the potential difference $E|e|d/t$, for $U/t=1$ and magnetic field $H/t=0$ (a) and $H/t=4$ (b).}
\end{center}
\end{figure}

\begin{figure}[h]
\begin{center}
\includegraphics[width=0.9\textwidth]{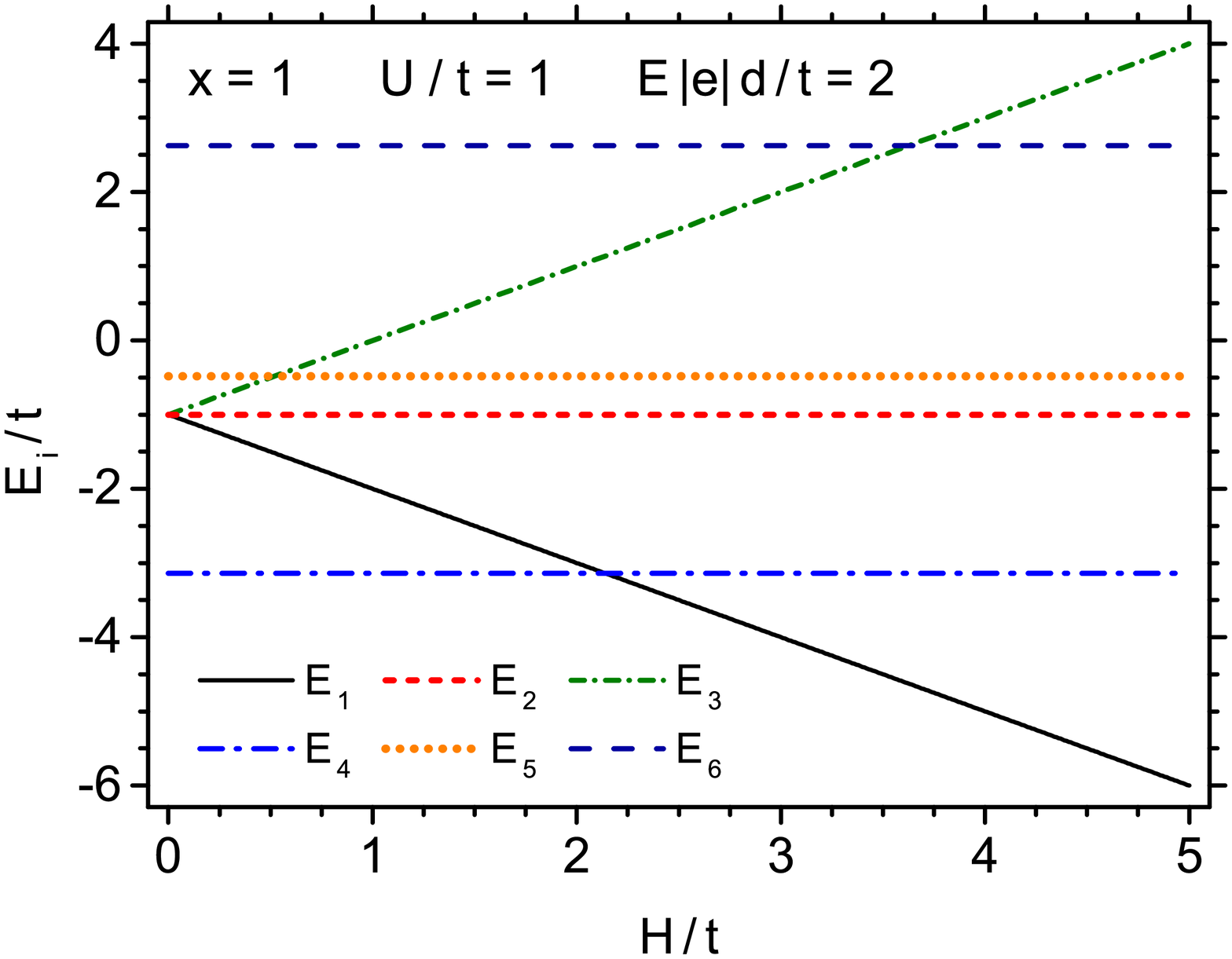}
\vspace{2mm}
\caption{\label{Fig17} The eigenenergies of the 6 quantum states corresponding to two electrons per dimer as a function of the magnetic field $H/t$, for $E|e|d/t=2$ and $U/t=1$.}
\end{center}
\end{figure}

For the case of $x=1$, i.e. half-filling of the energy states (in the presence of two electrons in a dimer) the chemical potential is constant and equal to $\mu=U/2$. There are 6 quantum states corresponding to the total occupation number equal to 2 for a dimer (out of 16 states in total). In the paper we make use of the grand canonical ensemble formalism, with the average number of electrons fixed with chemical potential. However, in the ground state the particle number fluctuations vanish and the behaviour of the system is only ruled by the states with exactly 2 electrons per dimer. Below we list the corresponding eigenenergies of the states.

The first three eigenenergies are equal to:
\begin{eqnarray}\label{eq:eigen1}
E_1&=&-U-H\\
E_2&=&-U\\
E_3&=&-U+H
\end{eqnarray}
and do not depend on the electric field $E$.

The remaining three eigenenergies $E_4,E_5,E_6$ do not depend on the magnetic field $H$ and constitute the roots of the following cubic equation:
\begin{equation}
E_i^3+UE_i^2-\left(4t^2+|e|^2 d^2 E^2\right)E_i-U |e|^2 d^2 E^2=0.
\end{equation}
These (real) roots are given by the following analytic formulas:
\begin{eqnarray}\label{eq:eigen2}
E_4&=&\frac{1}{3}\left[-\frac{\left(1-i \sqrt{3}\right) B}{2^{4/3}
}-\frac{\left(1+i \sqrt{3}\right) \left(12 t^2+U^2+3 |e|^2 d^2 E^2\right)}{2^{2/3} B}-U\right]\\
E_5&=&\frac{1}{3}\left[   -\frac{\left(1+i \sqrt{3}\right) B}{2^{4/3}
}-\frac{\left(1-i \sqrt{3}\right) \left(12 t^2+U^2+3 |e|^2 d^2 E^2\right)}{2^{2/3} B}-U\right]\\
E_6&=&\frac{1}{3}\left[\frac{B}{ 2^{1/3}}+\frac{2^{1/3} \left(12 t^2+U^2+3
   |e|^2 d^2 E^2\right)}{ B}-U\right],
\end{eqnarray}
where 
\begin{equation}
B=\left(4\sqrt{A}+36 t^2 U-2 U^3+18 U |e|^2 d^2 E^2\right)^{1/3}
\end{equation}
with
\begin{equation}
A=\left(18t^2U+U^3+9U |e|^2 d^2 E^2\right)^2-4\left(12t^2+U^2+3|e|^2 d^2 E^2\right)^3.
\end{equation}

In the absence of the external fields the eigenenergies reduce to:
\begin{eqnarray}
E_1&=-U\\
E_2&=-U\\
E_3&=-U\\
E_4&=&\frac{-U-\sqrt{16t^2+U^2}}{2}\\
E_5&=&0\\
E_6&=&\frac{-U+\sqrt{16t^2+U^2}}{2}.
\end{eqnarray}

The states labelled with $i=1,2,3$ correspond to spin triplet states with the total spin quantum number equal to $S=1$. For $i=1$ the spin projection quantum number $S^{z}$ is equal to 1, while for $i=2$ it amounts to 0 and for $i=3$ it takes the value of -1. The states labelled with $i=4,5,6$ are of singlet nature, with $S=0$ and $S^{z}=0$.

The behaviour of the individual energy states as a function of the external electric and magnetic field is the key factor shaping the response of the dimer to these fields. Therefore, it is instructive to analyse the dependence of the eigenenergies $E_i$ for $i=1,\dots,6$ on the electric and magnetic field. 

Fig.~\ref{Fig16} presents the evolution of the energy spectrum of the half-filled dimer with $U/t=1$ when the electric field is varied, in the absence of the magnetic field [Fig.~\ref{Fig16}(a)] or in the presence of noticeably high magnetic field $H/t$=4 [Fig.~\ref{Fig16}(b)]. It is clearly visible that for $H=0$ the ground state is always of singlet type. On the other hand, for $H/t=4$, below some critical electric field the triplet ground state is observed, whereas above the critical electric field a singlet-type ground state is restored. 

To complete the microscopic picture, Fig.~\ref{Fig17} shows the analogous dependence of the eigenenergies on the magnetic field, for $U/t=1$ and $E|e|d/t=2$. Below a critical magnetic field a singlet ground state is seen, while the increase in the magnetic field switches the system to a triplet state with $S=1$. 

It can be commented that in the cases discussed in Figs.~\ref{Fig16}(b) and \ref{Fig17}, the value of the critical field (either magnetic or electric one) results from the competition between the total energies of singlet state ($i=4$) and triplet state ($i=1$). As a consequence, it might be calculated by solving the equation $E_1=E_4$, where the energies are given by Eq.~\ref{eq:eigen1} and \ref{eq:eigen2}, respectively. Let us mention that the detailed discussion of the behaviour of the critical field as a function of model parameters was provided in our earlier work Ref.~\cite{Balcerzak2018} (see Fig.~1 therein). In both cases shown in Figs.~\ref{Fig16}(b) and \ref{Fig17} the transitions are of discontinuous type, involving a step-like change of the magnetic or electric properties in the ground state.


\bibliographystyle{elsarticle-num}

\end{document}